\documentclass[%
reprint,
superscriptaddress,
 amsmath,amssymb,
 aps,
]{revtex4-2}

\usepackage{graphicx}
\usepackage{dcolumn}
\usepackage{bm}
\usepackage{hyperref}
\usepackage{amsmath}
\usepackage{xcolor}

\begin{document}

\preprint{}

\title{Reconstruction of 2D line-integrated electron density using angular filter refractometry and a fast marching Eikonal solver}

\author{B. McCluskey}
    \email{bpmcclu2@pppl.gov}
\author{J. Griff-McMahon}%
\affiliation{Department of Astrophysical Sciences, Princeton University, Princeton, New Jersey 08544}
\affiliation{Princeton Plasma Physics Laboratory, Princeton, New Jersey 08540}

\author{D. Haberberger}
\affiliation{Laboratory for Laser Energetics,  Rochester, New York 14623}

\author{V. Valenzuela-Villaseca}
\affiliation{Department of Astrophysical Sciences, Princeton University, Princeton, New Jersey 08544}

\author{H. Landsberger}
\affiliation{Department of Astrophysical Sciences, Princeton University, Princeton, New Jersey 08544}
\affiliation{Princeton Plasma Physics Laboratory, Princeton, New Jersey 08540}

\author{W. Fox}
\affiliation{Princeton Plasma Physics Laboratory, Princeton, New Jersey 08540}
\affiliation{Department of Physics,  University of Maryland, College Park, Maryland 20742}

\date{\today}

\begin{abstract}
    Refraction of an optical probe beam by a plasma can be measured with angular filter refractometry (AFR), which produces an image containing intensity contours that correspond to curves of constant refraction angle. Further analysis is required to reconstruct the underlying line-integrated electron density. Most prior efforts to calculate density from AFR data have been limited to 1D analysis or forward-fitting techniques. In this paper, we detail the use of a fast marching Eikonal solver to directly invert AFR data and obtain the 2D line-integrated electron density. The analysis method is first verified with synthetic data and then applied to laser-driven experiments of single and double plume expansion collected at the OMEGA EP Laser Facility. The calculated densities agree with 1D results and are shown to be consistent with the original AFR measurements via forward modeling. We also discuss how additional measurements could improve the precision of this technique.
\end{abstract}

\maketitle


\section{\label{sec:Introduction} Introduction}

    Electron density is one of the key parameters characterizing laser-produced plasmas, and various optical diagnostics have been developed to measure it \cite{harilal, batani2019optical}. Angular filter refractometry (AFR) is a technique that uses refractive measurements to infer the line-integrated electron density over large spatial areas (${\sim}4$ mm$^2$) \cite{haberberger_afr}. AFR provides a method to quantitatively explore density regimes ($10^{20}$-$10^{21}$ cm$^{-3}$) that are relevant to high-energy-density physics but inaccessible by other diagnostics \cite{haberberger_afr, angland}.

    AFR has been utilized in a wide range of experiments, including studies of collisionless shocks \cite{schaeffer}, laser-driven particle acceleration \cite{kordell}, plasma channeling \cite{ivancic2015pre, Murakami_2016}, and inertial confinement fusion \cite{haberberger-inner-shell}. Most prior efforts to reconstruct the density from raw AFR data rely on either calculating the density in a quasi-1D region of the plasma \cite{Li, fox, Peebles_2017} or assuming an analytic model for the density and iteratively adjusting the parameters until the resulting AFR signal closely matches the experimental measurements \cite{angland, bailly-grandvaux, ucsd-thesis, Murakami_2016}. These methods have limited applicability since they are either symmetry or model dependent, which motivates the development of a direct analysis technique. This need is further underscored by recent improvements in AFR filter design that have advanced the diagnostic's effectiveness \cite{heuer-filters}.

    In this paper, we develop and present an analysis technique that directly calculates the 2D line-integrated electron density from AFR measurements by numerically solving the Eikonal equation. This approach does not require an axisymmetric plasma or an assumed density profile. \textcite{ivancic2015thesis} was the first to analyze AFR data with an Eikonal solver, and we expand on his work by validating the technique with synthetic data, considering non-zero boundary conditions, and applying the method to non-axisymmetric plasmas. This paper is organized as follows. In Section \ref{sec:Overview of angular filter refractometry}, we summarize the theory behind AFR. In Section \ref{sec:Analysis}, we present the three components of the analysis method: the numeric Eikonal solver, the interpolation scheme, and the boundary conditions. In Section \ref{sec:Application}, we apply the technique to both synthetic data and experimental measurements obtained at the OMEGA EP Laser Facility. Section \ref{sec:Conclusion} discusses improvements that could be implemented to address the limitations of the technique. 

\section{\label{sec:Overview of angular filter refractometry} Overview of angular filter refractometry}

    AFR was developed by \textcite{haberberger_afr} to measure line-integrated electron density. An optical probe beam propagating along the $z$-axis passes through a plasma. Due to the refractive index of the plasma, the probe accumulates a phase shift relative to vacuum propagation that is given by
    \begin{equation}\label{eq:phase}
        \phi(x,y) = -\frac{\pi}{\lambda_p n_{cr}}\int_{-\infty}^{\infty} n_e(x,y,z) \,dz,
    \end{equation}
    where $\lambda_p$ is the probe wavelength and $n_{cr} = [4\pi^2 \epsilon_0 m_ec^2 / e^2 ]\lambda_p^{-2}$ is the critical density \cite{hutchinson2002principles}. Equation \eqref{eq:phase} assumes the refractive index for an unmagnetized plasma, $\sqrt{1-n_e/n_{cr}}$, and $n_e \ll n_{cr}$. If the probe beam develops a spatially non-uniform phase across its wavefront, it will refract by the angle 
    \begin{equation}\label{eq:refracted angle}
        \bm\theta (x,y) = \frac{\lambda_p}{2\pi} \nabla\phi(x,y)
    \end{equation}
    relative to the optical axis (i.e., the $z$-axis) \cite{hutchinson2002principles}. 

    \begin{figure*}[!t]
        \centering
        \includegraphics[width=\textwidth]{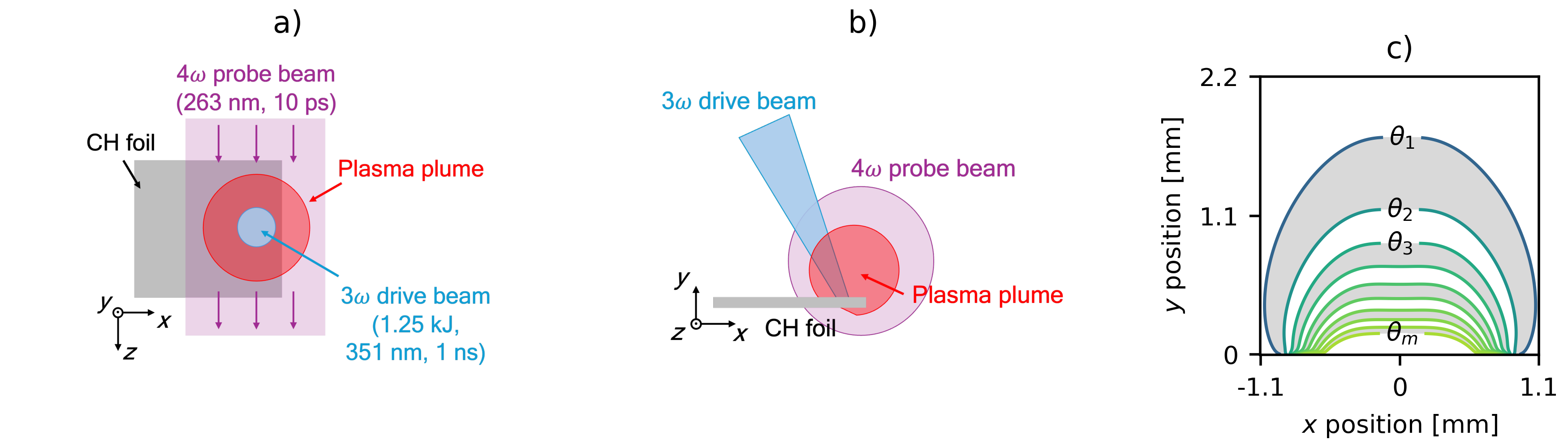}
        \caption{Top-down (a) and side-on (b) schematics for experiments studying laser-driven expansion of a single plasma plume from a planar target.  Single plumes were generated by a 3$\omega$ drive beam (1.25 kJ, 1 ns, 351 nm) with a peak intensity of $3\times10^{14}$ W/cm$^{2}$ incident on a 25 $\mu$m thick plastic (CH) target. AFR was performed along the $z$-axis via a 4$\omega$ probe beam (10 mJ, 10 ps, 263 nm). All experiments \cite{fox} were performed at the OMEGA EP Laser Facility. (c) A synthetic AFR image for a single plume with several contours of constant $\theta$ indicated. The solid gray bands represent regions of high intensity. We enumerate the contours so that the outermost edge is the 1st contour and the innermost edge is the $m$th contour. The density profile is generated using the model presented in \textcite{angland}, with parameters $n_0=1.375\times10^{21}$ cm$^{-3}$, $A=0.8605$, $L_{y1}=0.4289$ mm, $L_{y2}=0.08823$ mm, $L_{xz}= 0.583$ mm, $c_1=1.558$, $c_2=0.4671$, $c_3 = 0.4036$ mm$^{-1}$, and $\lambda_p=263$ nm.}
        \label{fig:single_plume_schematic}
    \end{figure*}

    The essential feature of AFR is an angular filter (alternating transparent and opaque concentric rings) located in the Fourier plane of the imaging optics. At this location, the beam's radial distance from the optical axis is directly proportional to $|\bm\theta|=\theta$, with the constant of proportionality determined by the experimental geometry. Therefore, the filter either blocks or transmits rays depending on $\theta$, with transitions occurring at a set of fixed angles determined by the radii of the concentric rings. The resulting image consists of bright and dark bands whose boundaries represent contours of constant $\theta$ \cite{haberberger_afr}. Each contour is labeled with one $\theta$ from the set of possible transition angles \cite{heuer-filters}. Since $\theta$ is only measured along band edges, AFR measurements are inherently sparce, and the spatial resolution is limited by the local band width. Figure \ref{fig:single_plume_schematic} shows an example experimental schematic and a synthetic AFR image, which demonstrates the band structure produced by an expanding plasma plume and the location of the $\theta$ contours. 

\section{\label{sec:Analysis} Analysis method}

    The goal of AFR analysis is to take an input measurement of $\theta$ and calculate $\phi$ ($\propto\int n_e dz$). These two quantities are related through the magnitude of Eq.~\eqref{eq:refracted angle}:
    \begin{equation}\label{eq:Eikonal}
        |\nabla\phi(\mathbf{x})|= \frac{2\pi}{\lambda_p}\theta(\mathbf{x})\text{ for } \mathbf{x}\in\Omega,
    \end{equation}
    where $\mathbf{x}=(x,y)$ is a point in the AFR measurement domain $\Omega\subset \mathbb{R}^2$ and $\theta(\mathbf{x})>0$. Equation \eqref{eq:Eikonal} takes the form of the Eikonal equation, which has been extensively studied due to its wide applications in physics, control theory, and optimization \cite{sethian-level_set_method}. Although nonlinear, the Eikonal equation is a first-order partial differential equation that can be numerically solved given sufficient boundary conditions on $\partial\Omega\subset\Omega$. Therefore, 2D AFR analysis requires a numeric solver, an interpolation scheme for sparse $\theta$ data, and a boundary condition for $\phi$.
    
\subsection{Numeric solver}

    The fast marching method (FMM) developed by \textcite{sethian-level_set_method} is a common method to numerically solve the Eikonal equation. Special discretization schemes for the gradient operator in Eq.~\eqref{eq:Eikonal} must be employed to ensure an entropy-satisfying solution (i.e., a single-valued $\phi$ solution) \cite{white}. For 2D Cartesian grids with regular spacing $\Delta x$ and $\Delta y$ in the $x$ and $y$ directions, a common approach is
    \begin{equation}\label{eq:Eikonal_discretization}
        |\nabla\phi|^2 \equiv \sum_{\alpha=x,y} \text{max}\left(D^{-\alpha}_{ij} \phi, - D^{+\alpha}_{ij} \phi, 0 \right)^2,
    \end{equation}
    where $i$ and $j$ are integers labeling each grid point, and $D$ is a first-order differential operator defined as 
    \begin{equation}
        D_{ij}^{\pm x} \phi = \frac{\phi_{i\pm1,j}-\phi_{ij}}{\pm \Delta x} ~~~~~~~ D_{ij}^{\pm y} \phi = \frac{\phi_{i,j\pm1}-\phi_{ij}}{\pm \Delta y}
    \end{equation}
    with $\phi_{ij} \equiv \phi(i \Delta x, j\Delta y)$ \cite{Rouy}. This discretization transforms Eq.~\eqref{eq:Eikonal} into a system of quadratic equations for $\phi_{ij}$, which can be iteratively solved. Since Eq.~\eqref{eq:Eikonal_discretization} relies on upwind differences, $\phi_{ij}$ can only be affected by neighbors with smaller values of $\phi$ \cite{sethian-review}.

    The FMM takes advantage of the upwind discretization to build the solution outward from the smallest value of $\phi$ \cite{sethian-review}. To avoid unnecessary steps, the FMM restricts attention to the narrow band of \textit{trial} points that directly neighbor \textit{known} points. All other points are labeled \textit{unknown} and excluded from consideration. The solver begins by calculating $\phi$ at each trial point. Upon completion, the point with the smallest $\phi$ is relabeled as known since it cannot be affected by any of the other trial points. The trial region is then expanded to include adjacent points that were previously labeled unknown. These steps are repeated until all points in the domain are known. A detailed discussion of the FMM algorithm can be found in other publications \cite{sethian-review, sethian-level_set_method, white}. 

\subsection{Data interpolation}

    Implementing the FMM solver requires knowledge of the source term $\theta(\mathbf{x})$ at all $\mathbf{x}$ in the domain $\Omega$; however, AFR measurements are inherently sparse, so an interpolation is required to evaluate $\theta$ in the region between contours. For plasmas generated by laser-solid interactions, we assume that the density profile is roughly exponential in the direction outward from the laser spot \cite{fox}. Therefore, we linearly interpolate $\ln(\theta)$. This `exponential interpolation' restricts the domain $\Omega$ to the points between the inner and outermost contours. In Section \ref{sec:Application}, we discuss complications that arise when applying this method to double plasma plume geometries. Methods to improve the interpolation are proposed in Section \ref{sec:Conclusion}.

\subsection{Boundary conditions}

    The final component needed to numerically solve the Eikonal equation is an adequate boundary condition for $\phi$ along $\partial\Omega$. Although this information could be obtained directly from separate interferometry data (see the discussion in Section \ref{sec:Conclusion}), such a setup was not available for the experiments presented in this paper. Therefore, we must develop an approximate boundary condition from the AFR data alone. Based on the characteristics of the Eikonal equation, information propagates ``up'' the density profile starting from the lowest value of $\phi$ on the boundary \cite{zhao2005fast}. Therefore, only the low-$\phi$ boundary conditions are required. For plasmas generated by laser-solid interaction, the smallest values of $\phi$ occur at the outermost contour, which is the set of points $\lbrace\mathbf{x}_1\rbrace$ such that $\theta(\mathbf{x}_1)=\theta_1$. Here, we seek to determine the boundary condition $\phi(\mathbf{x}_1)$ for all points along the $\theta_1$ (outermost) contour. 

    We start by assuming that $\phi$ can be locally described as an exponential near the $\theta_1$ contour:
    \begin{equation}\label{eq:ansatz}
        \phi(\mathbf{x}) = \phi_0+\phi_1(\mathbf{x})\exp\left( \frac{-\hat{\mathbf{g}}(\mathbf{x})\cdot (\mathbf{x}-\mathbf{x}_1)}{L(\mathbf{x})} \right).
    \end{equation}
    Here, $\hat{\mathbf{g}}$ indicates the direction of steepest descent, which is roughly normal to the $\theta_1$ contour, and $L$ is the gradient scale length. In this construction, we assume that $\phi_1$, $\mathbf{\hat{g}}$, and $L$ vary in space, but that the variation is weak compared to the spatial dependence of the exponential term. Any constant phase offset $\phi_0$ due to a uniform background density remains undetected by AFR since the diagnostic is only sensitive to $\nabla\phi$.

    Taking derivatives of $\phi(\mathbf{x})$, we obtain
    \begin{equation}\label{eq:theta_ansatz}
        \theta(\mathbf{x}) =  \frac{\lambda_p}{2\pi} |\nabla \phi(\mathbf{x})| \\
        \approx \frac{\lambda_p}{2\pi} \frac{\phi_1}{L} \exp\left( \frac{-\hat{\mathbf{g}}\cdot (\mathbf{x-x}_1)}{L} \right)
    \end{equation}
    and 
    \begin{equation}
        |\nabla \theta(\mathbf{x})| \approx \frac{\theta(\mathbf{x})}{L},
    \end{equation}
    where we have ignored derivatives of $\phi_1$, $\mathbf{\hat{g}}$, and $L$ compared to the exponential. For all points $\lbrace\mathbf{x}_1\rbrace$ along the contour, the refraction angle $\theta(\mathbf{x}_1)=\theta_1$ is constant, and $|\nabla\theta(\mathbf{x}_1)|$ is directly available from the interpolated data. We evaluate the last two equations at position $\mathbf{x}_1$, eliminate $L$, and solve for $\phi_1(\mathbf{x}_1)$. Substituting this result into Eq.~\eqref{eq:ansatz}, we arrive at
    \begin{equation}\label{eq:bc}
        \phi(\mathbf{x}_1)=\phi_0+\frac{2\pi}{\lambda_p}\frac{\theta_1^2}{|\nabla\theta(\mathbf{x}_1)|}.
    \end{equation}
    This provides a boundary condition along the $\theta_1$ contour that is completely derivable from the AFR data. Synthetic numerical examples below show that this boundary condition is accurate to ${\sim}30$\% for typical plasma profiles when $\theta$ is perfectly interpolated.     
    
    Equation \eqref{eq:bc} represents the boundary condition for a plasma in which the density drops exponentially in the direction normal to the outermost band. To the extent that these conditions are not fulfilled, Eq.~\eqref{eq:bc} will not be an accurate boundary condition and will result in proportional offsets from the true boundary value. Nevertheless, such inaccuracies are mitigated by the fact that linear offsets become small as the FMM solver marches up the exponentially increasing density profile. Thus, most error in the boundary condition is confined near the outermost contour. 

\section{\label{sec:Application} Application}

    AFR has been primarily used to probe laser-solid interactions, in which drive beams irradiate solid targets, ablate material, and generate plumes of plasma that expand from the target surface \cite{haberberger_afr}. These experiments typically investigate the dynamics of a single plume (e.g., expansion rate \cite{haberberger_afr}, heat transport \cite{hawreliak2004thomson}, and self-magnetization \cite{stamper1978studies}) or the interaction of multiple colliding plumes (relevant to magnetic reconnection \cite{nilson-prl} and magnetic flux compression \cite{sladkov2024saturation}). Here, we apply the analysis method developed in Section \ref{sec:Analysis} to calculate $\int n_edz$ for both single and double plumes expanding from a planar target. For each case, we first verify the analysis procedure with synthetic data before applying it to experimental measurements \cite{fox} collected at the OMEGA EP Laser Facility.

\subsection{Single plumes}

    We begin by analyzing synthetic single plume AFR data. The geometry of single plume experiments is shown schematically in Figs.~\ref{fig:single_plume_schematic}a and \ref{fig:single_plume_schematic}b. A synthetic density profile is generated using the analytic model presented in \textcite{angland}, which treats the density as a decaying exponential in the target-normal direction ($y$-axis) and a super-Gaussian in the radial direction ($x$-$z$ plane). The resulting synthetic AFR image is shown in Fig.~\ref{fig:single_plume_schematic}c, along with several contours of constant $\theta$.

\begin{figure}
        \centering
        \includegraphics[width=\linewidth]{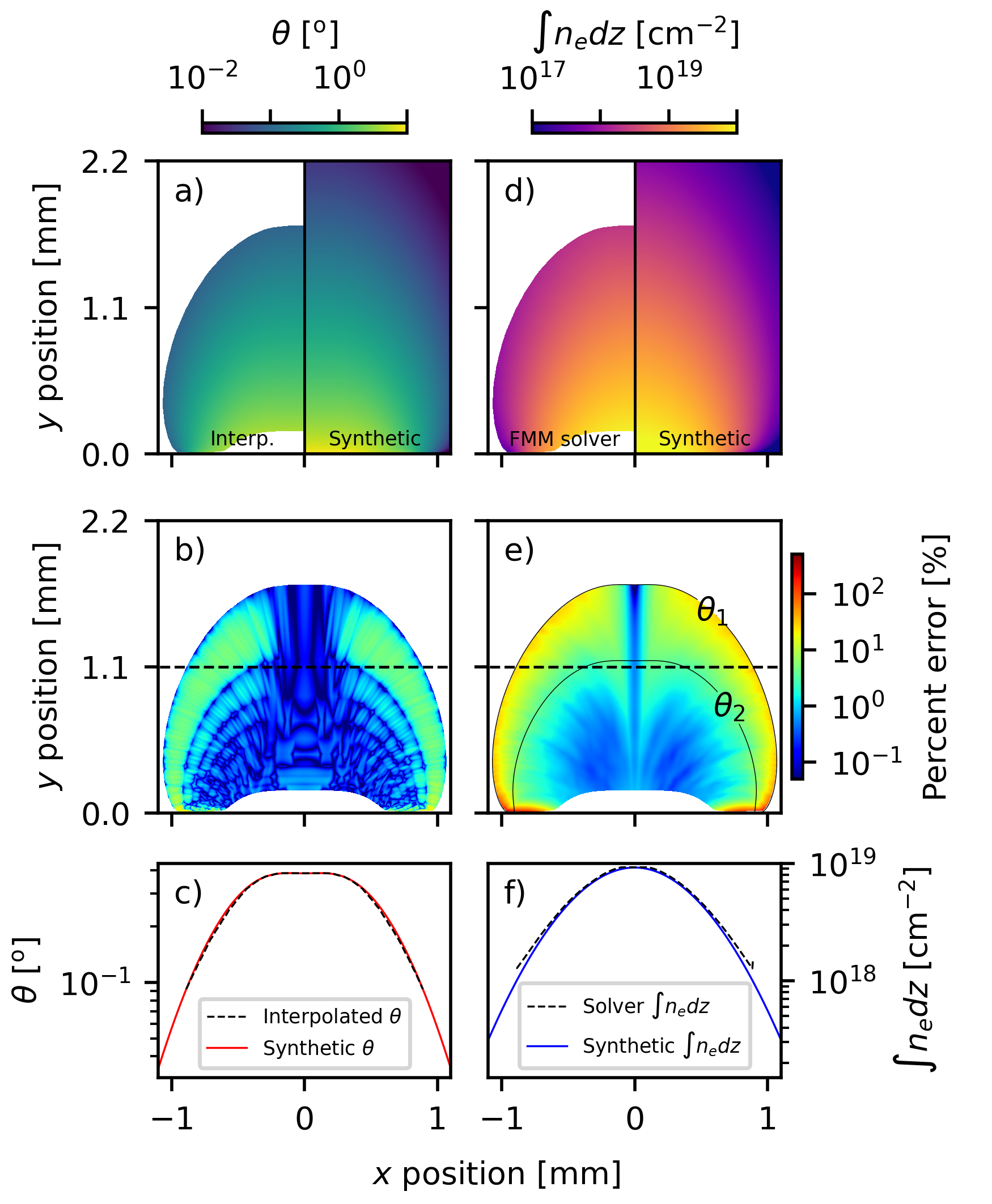}
        \caption{(a) Comparison of the exponentially interpolated (left) and synthetic (right) $\theta$ profiles for a single plume. (b) The $\theta$ interpolation error. (c) Lineouts at $y=1.1$ mm of the interpolated and synthetic $\theta$. (d) The reconstructed $\int n_edz$ profile calculated by the FMM solver (left) compared to the synthetic density profile (right). The solver uses Eq.~\eqref{eq:bc} as the boundary condition. (e) The FMM solution error. The $\theta_1$ and $\theta_2$ contours are shown for reference. (f) Lineouts at $y=1.1$ mm of the FMM solver and synthetic $\int n_edz$.}
        \label{fig:single_synthetic_error}
    \end{figure}

    The $\theta$ contours are the input to the AFR analysis scheme. First, $\theta$ is exponentially interpolated. Fig.~\ref{fig:single_synthetic_error}a shows the interpolated and synthetic $\theta$ side by side. Coarse spatial measurements of $\theta$ combined with non-exponential terms in the synthetic density profile lead to some error in the interpolation (see Fig.~\ref{fig:single_synthetic_error}b). The typical error is $<10\%$ and is mainly confined to regions where the band spacing is the largest and where the super-Gaussian term in the synthetic profile dominates. A comparison of $\theta$ lineouts in Fig.~\ref{fig:single_synthetic_error}c further demonstrates that the interpolation closely replicates the synthetic profiles.

    \begin{figure*}
        \centering
        \includegraphics[width=\textwidth]{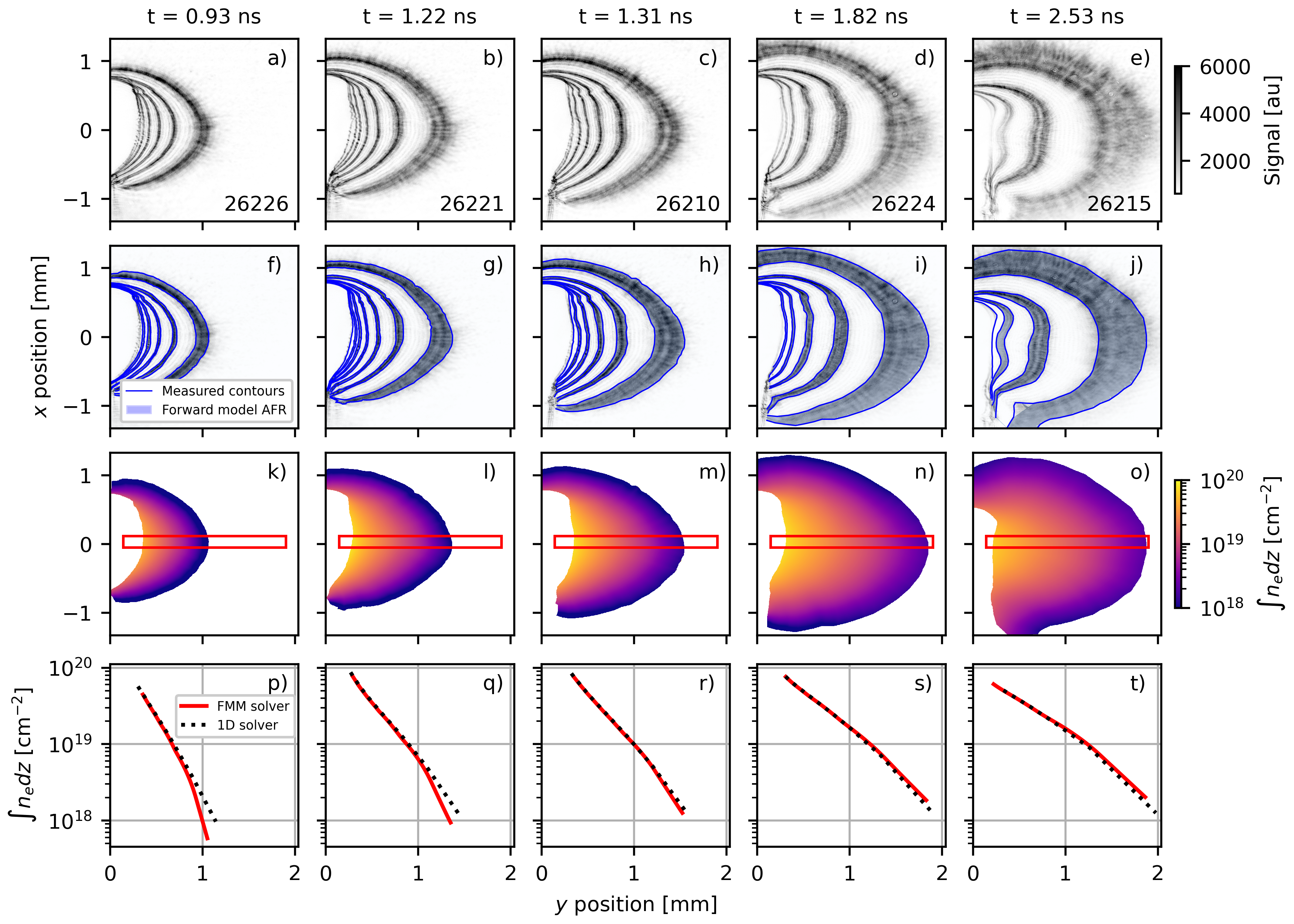}
        \caption{(a)-(e) The experimental AFR images at time $t$ after the drive beam is incident on the target. Each frame represents a separate experimental shot. (f)-(j) The measured contours (blue lines) and the forward-modeled AFR bands (blue shading) overlaid on the original AFR images. (k)-(o) The FMM reconstruction of $\int n_edz$. (p)-(t) Lineouts taken from the red rectangle in subplots (k)-(o) are compared with the results of the 1D solver used by \textcite{fox}.}
        \label{fig:single_plume}
    \end{figure*}
    
    Having verified the $\theta$ interpolation scheme, we apply the FMM solver. Equation \eqref{eq:bc} is employed as the boundary condition, with $\phi_0=0$ since we expect negligible plasma density far from the solid target. The FMM $\int n_edz$ solution is compared to the synthetic profile in Fig.~\ref{fig:single_synthetic_error}d. The solution error, shown in Fig.~\ref{fig:single_synthetic_error}e, reaches approximately ${\sim}30\%$ near the $\theta_1$ contour. It is not surprising that Eq.~\eqref{eq:bc} introduces error near the boundary since the density profile includes a super-Gaussian term and is not purely exponential. Nevertheless, errors at the boundary become insignificant ($<5\%$) as the density exponentially increases toward the laser footprint (see Fig.~\ref{fig:single_synthetic_error}f). 

    Given the effectiveness of the FMM on synthetic data, we now apply it to experimental data (see Figs.~\ref{fig:single_plume_schematic}a and \ref{fig:single_plume_schematic}b for experimental details). Raw AFR images taken at various times are shown in Figs.~\ref{fig:single_plume}a-e. The high-contrast, low-frequency structures are the AFR bands, while the high-frequency features are diffractive effects caused by the sharp edges of the angular filter. The large regions of low signal near the origin are caused by steep density gradients near the laser footprint that refract the probe beam outside of the collection optics (${\sim}7^{\circ}$) \cite{heuer-filters}. The AFR analysis begins by manually locating the contours, which are shown as blue lines in Figs.~\ref{fig:single_plume}f-j. The mapping from contour to refraction angle is performed using an OMEGA EP facility calibration measurement \cite{haberberger_afr}. Next, the boundary conditions are approximated with Eq.~\eqref{eq:bc}, and the FMM solver produces the density profiles shown in Figs.~\ref{fig:single_plume}k-o. Lineouts of the FMM solution along the $y$ axis agree well with 1D solvers \cite{fox}, as seen in Figs.~\ref{fig:single_plume}p-t. Differences between the FMM and 1D solutions are attributed to discrepancies in how each solver locates the $\theta$ contours. 

    From the reconstructed density, we forward model AFR bands and compare them with the original experimental measurements. In Figs.~\ref{fig:single_plume}f-j, we observe excellent agreement between the forward model and the experimental data, demonstrating that the FMM solution is consistent with the original measurement. The forward AFR model assumes that the probe beam is uniform; however, there is evidence for aberrations in the experimental beam \cite{angland}.
    
\subsection{Double plumes}

    \begin{figure*}[ht!]
        \centering
        \includegraphics[width=\textwidth]{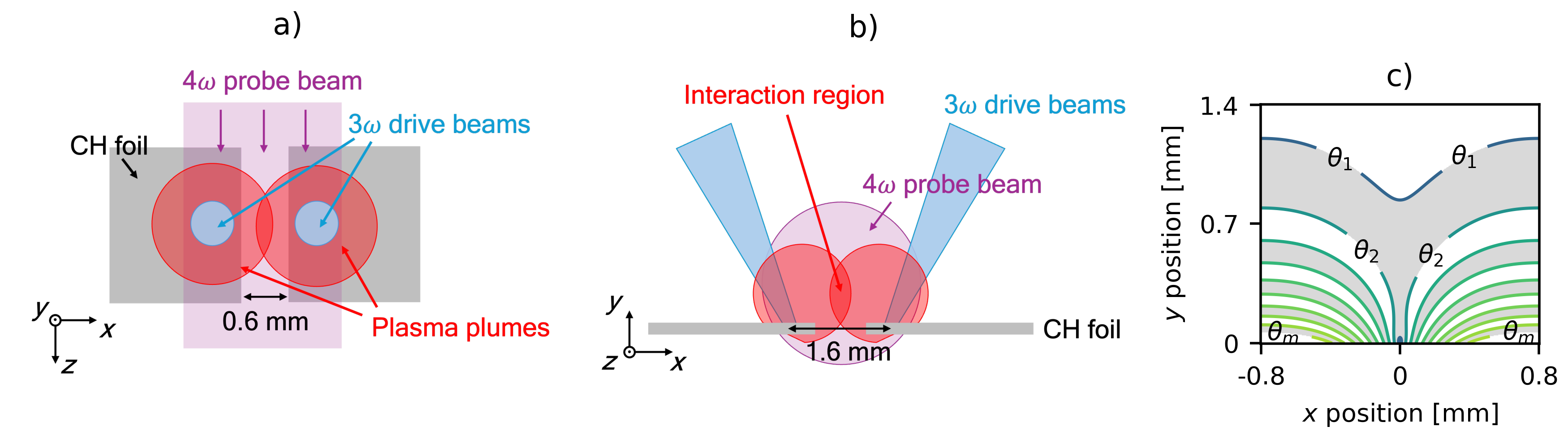}
        \caption{Top-down (a) and side-on (b) schematics for the double plume experiments at the OMEGA EP Laser Facility. Two identical CH targets are placed 0.6 mm apart, with the laser foci separated by 1.6 mm. Drive beams arrive at each target simultaneously and produce two plumes that expand and interact. Laser parameters for the drive and probe beams are the same as the single plume experiment \cite{fox}. (c) A synthetic AFR image for a double plume experiment with several $\theta$ contours indicated. The synthetic density profile is constructed from the sum of two single plumes (see \textcite{angland}) that are separated by 1.6 mm in the $x$ direction and stretched by a factor of 1.4 in the $y$ direction. Notice the branch region near $x=0$ mm, where the AFR band has two edges that are both $\theta_2$ contours.}
        \label{fig:double_plume_schematic}
    \end{figure*}

    In double plume experiments, two beams are incident on adjacent targets, producing two plumes that expand and interact with each other (see Figs.~\ref{fig:double_plume_schematic}a and \ref{fig:double_plume_schematic}b). The double plume geometry, which is non-axisymmetric, complicates the interpolation scheme and boundary condition for AFR analysis. To understand these challenges, we first apply the FMM solver to the synthetic double plume data that is shown in Fig.~\ref{fig:double_plume_schematic}c. 
    
\begin{figure}
        \centering
        \includegraphics[width=\linewidth]{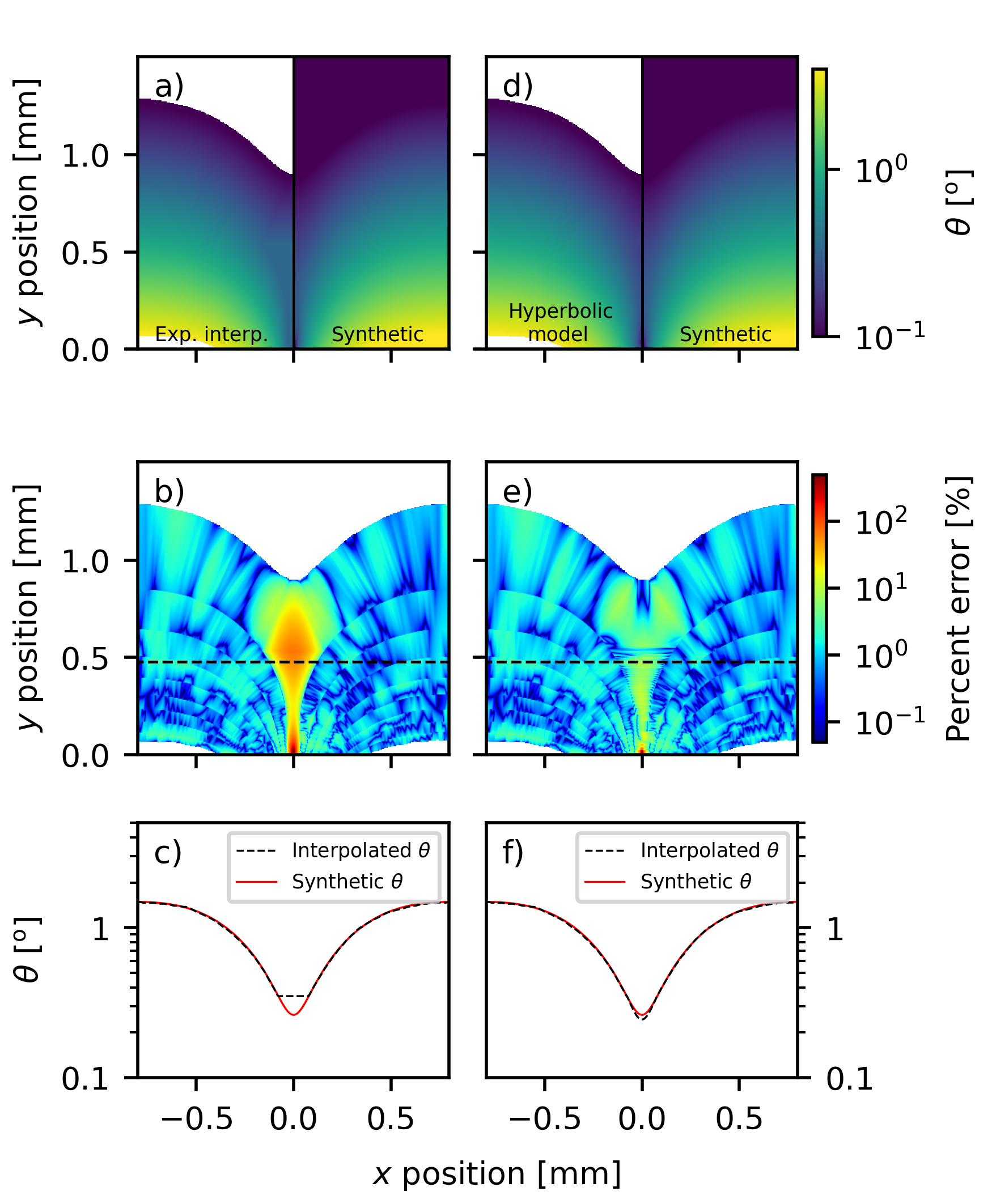}
        \caption{(a) Exponentially interpolated $\theta$ (left) compared to the synthetic double plume profile (right), and (b) the resulting interpolation error. (c) Lineouts at $y=0.48$ mm compare the exponential interpolation and the synthetic $\theta$. The interpolation predicts a constant $\theta$ in the branch region. (d) Interpolated $\theta$ using the hyperbolic model in the branch region (left) compared to the synthetic profile (right). (e) The hyperbolic model error. (e) Lineouts at $y=0.48$ mm compare the hyperbolic model and synthetic $\theta$. The hyperbolic model improves the prediction of $\theta$ in the branch region.}
        \label{fig:double_interpolation_error}
    \end{figure}

    It is necessary to develop a more sophisticated interpolation scheme to properly handle the plume interaction region. In the synthetic data shown in Fig.~\ref{fig:double_plume_schematic}c, the outermost AFR band contains a branch near $x=0$ mm where both band edges correspond to $\theta_2$ contours. Therefore, exponential interpolation incorrectly predicts a constant value of $\theta=\theta_2$ over the entire branch region (see Fig.~\ref{fig:double_interpolation_error}a), leading to significant error (see Fig.~\ref{fig:double_interpolation_error}b). Lineouts in Fig.~\ref{fig:double_interpolation_error}c confirm that the interpolation scheme incorrectly predicts constant $\theta$ across the branch. To address this challenge, we must model $\theta$ in the branch region. Assuming that $\theta$ reaches a local minimum near the center of the branch and that $\partial\theta/\partial x\gg\partial\theta/\partial y$, we can model $\theta$ as an asymmetric hyperbola
    \begin{equation}
        \theta(x,y)=
        \begin{cases} 
            \sqrt{\theta_\text{min}(y)+\left[m_-(x-x_\text{mid}(y))\right]^{2}} & x < x_\text{mid}(y) \\
            \sqrt{\theta_\text{min}(y)+\left[m_+(x-x_\text{mid}(y))\right]^{2}} & x \ge x_\text{mid}(y)
        \end{cases}
    \end{equation}
    where $\theta_\text{min}$, $x_\text{mid}$, $m_-$, and $m_+$ are analytically determined at each $y$ to enforce continuity of $\theta$ and $\partial\theta/\partial x$ at both $\theta_2$ contours. Note that $\theta$ and $\partial\theta/\partial x$ are continuous at $x=x_\text{mid}$ by construction. The hyperbolic model leads to a $\theta$ profile that closely resembles the synthetic data (see Fig.~\ref{fig:double_interpolation_error}d) and reduces the interpolation error (see Fig.~\ref{fig:double_interpolation_error}e). Lineouts in Fig.~\ref{fig:double_interpolation_error}f demonstrate that the hyperbolic model improves the prediction of $\theta$ in the branch region.
    
\begin{figure}
        \centering
        \includegraphics[width=\linewidth]{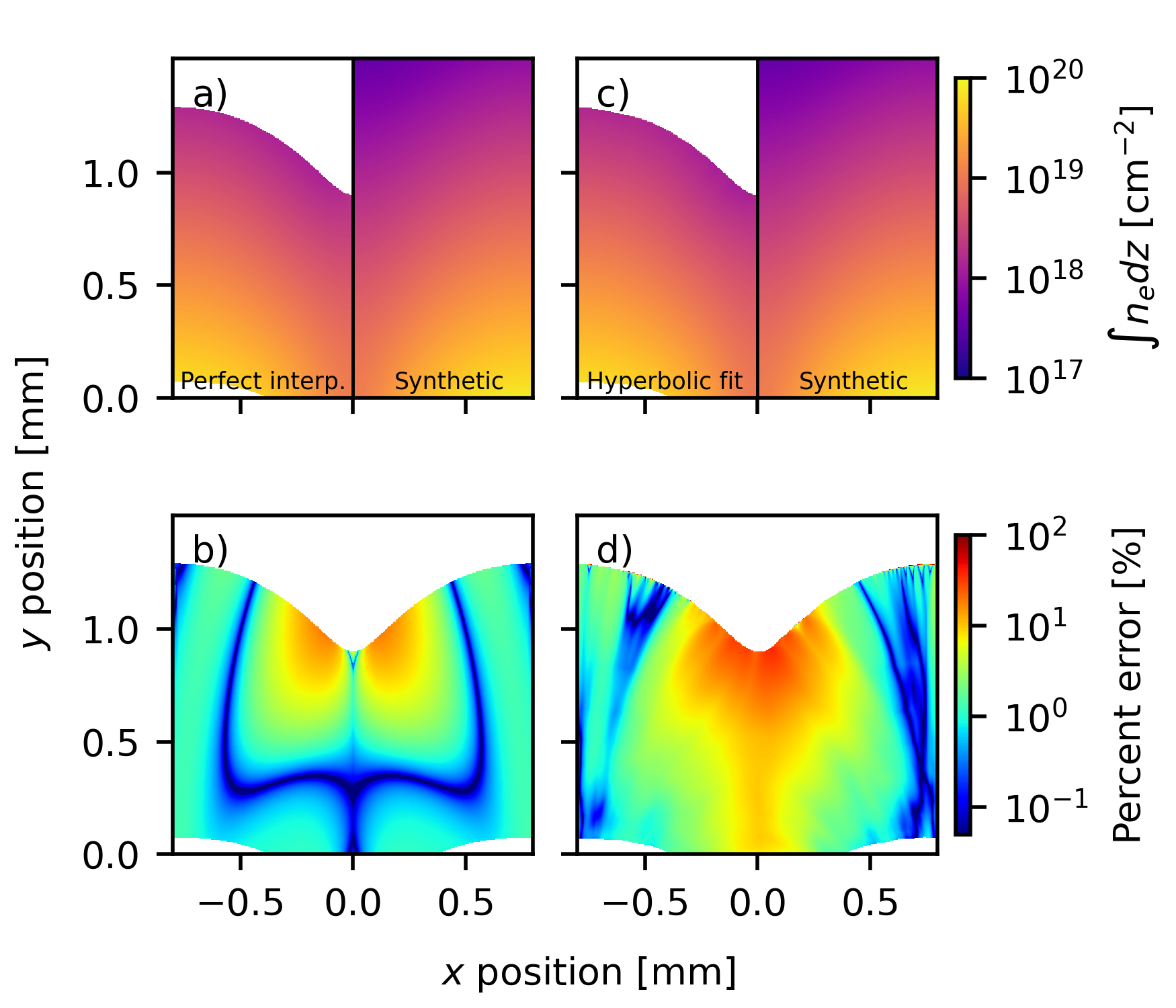}
        \caption{(a) The $\int n_edz$ profile reconstructed by the FMM solver when using perfectly interpolated $\theta$ and Eq.~\eqref{eq:bc} as the boundary condition (left) compared to the synthetic density profile (right). (b) The error in the FMM reconstruction due solely to the boundary condition. (c) The $\int n_edz$ profile reconstructed by the FMM solver when $\theta$ is interpolated (left) compared to the synthetic density profile (right). (d) The error in the FMM reconstruction when interpolation (including the hyperbolic model in the branch region) is included.}
        \label{fig:double_ndl_error}
    \end{figure}

    Non-axisymmetric plasmas also present a challenge for the boundary condition. Recall that  Eq.~\eqref{eq:bc} assumed that the density decreases exponentially in the radial direction. This is a weak assumption for double plume geometries. Nevertheless, when the effect of the boundary condition on the FMM solution is isolated by using perfectly interpolated $\theta$, we find that Eq.~\eqref{eq:bc} performs well for the double plume case (see Figs.~\ref{fig:double_ndl_error}a and \ref{fig:double_ndl_error}b). When the effects of interpolation are included, we find that the error in the interaction region ${\sim}10\%$, while the error near the boundary can approach 50-100$\%$ (see Figs.~\ref{fig:double_ndl_error}c and \ref{fig:double_ndl_error}d). Methods to obtain high fidelity boundary conditions for non-axisymmetric plasmas are discussed in Section \ref{sec:Conclusion}.

\begin{figure*}[t!]
        \centering
        \includegraphics[width=\textwidth]{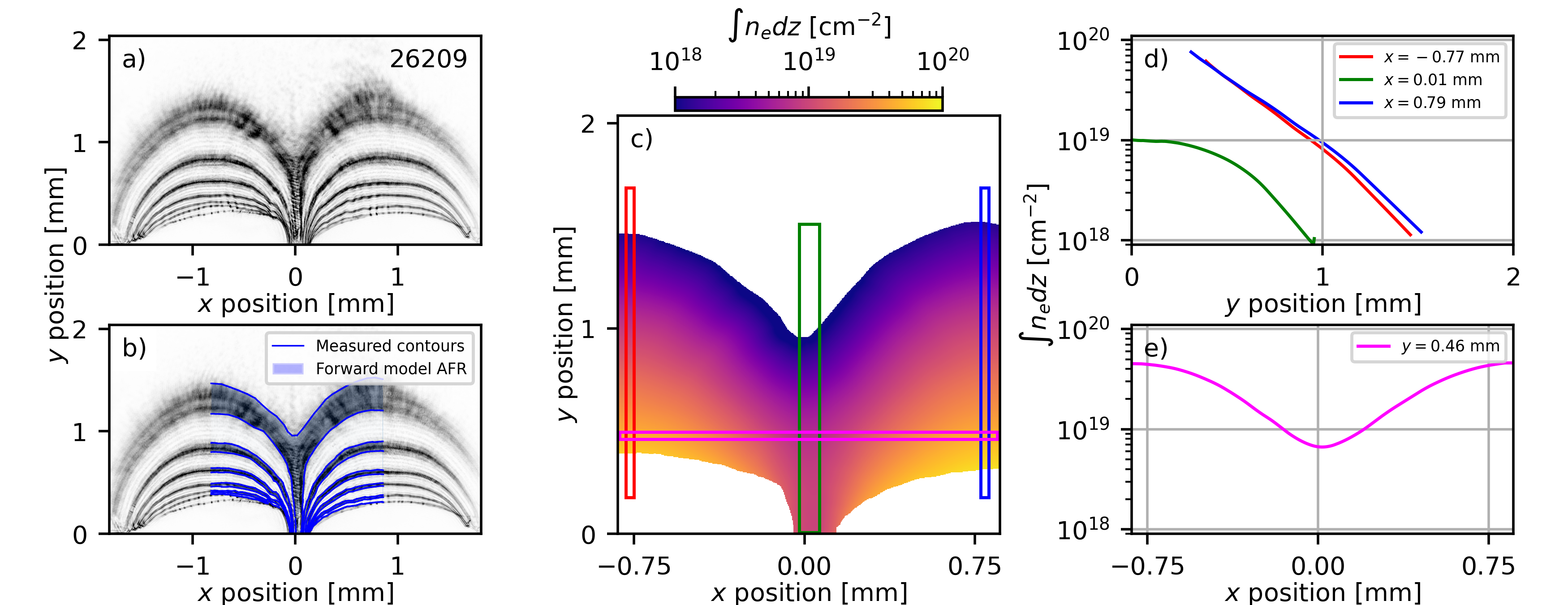}
        \caption{(a) Experimental double plume data at $t=1.2$ ns. (b) The manually located contours (blue lines) and the forward-modeled AFR bands (blue shading) overlaid on the experimental data. (c) The FMM $\int n_e dz$ solution for the experimental double plume data. This solution used the hyperbolic model of $\theta$ in the branch region and Eq. \eqref{eq:bc} for the boundary condition. Colored rectangles represent lineouts along the $y$ (d) and $x$ (e) axis. }
        \label{fig:double_experiment_ndl}
    \end{figure*}

    Having addressed the challenges of double plume analysis and verified our method with synthetic data, we now turn to experimental data. In Fig.~\ref{fig:double_experiment_ndl}a, we present the raw AFR image of two colliding plumes 1.2 ns after the drive beams are incident on the target. The bands are manually located (see Fig.~\ref{fig:double_experiment_ndl}b). Outside of the branch region, $\theta$ is exponentially interpolated, while the hyperbolic model is used inside the branch. Using Eq.~\eqref{eq:bc} as the boundary condition, the FMM reconstructs the $\int n_edz$ profile shown in Fig.~\ref{fig:double_experiment_ndl}c. A comparison of the forward-modeled AFR bands with the original measured contours in Fig.~\ref{fig:double_experiment_ndl}b demonstrates that the calculated density is consistent and that our analysis technique successfully analyzed this non-axisymmetric case. Furthermore, lineouts along the target normal (blue and red curves in Fig.~\ref{fig:double_experiment_ndl}d) agree with 1D solver results (not shown). Lineouts along (green curve in Fig.~\ref{fig:double_experiment_ndl}d) and across (Fig.~\ref{fig:double_experiment_ndl}e) the plume interaction region are also shown. 
    
    Our analysis neglects the high frequency structures that appear in the plume interaction region, as it is not possible to associate these features with a specific filter ring; we can only reconstruct density structures whose size is larger than the local band width. Simultaneous shadowgraphy images suggest that these structures correspond to large second derivatives in the plasma density.

\section{\label{sec:Conclusion} Conclusion}

    In this paper, we apply a fast marching method (FMM) Eikonal solver to directly reconstruct the 2D line-integrated electron density from angular filter refractometry (AFR) measurements. An interpolation scheme for sparse $\theta$ data and a boundary condition for $\int n_edz$ are developed and verified with synthetic data. We then apply the technique to analyze single and double plume experiments performed at the OMEGA EP Laser Facility. The resulting $\int n_edz$ profiles agree with 1D solvers and are consistent with the original AFR measurements. 
    
    The analysis method presented here is quite general and can be applied to many different plasma geometries; however, the details of the interpolation and boundary condition may have to be adjusted depending on the application. Furthermore, the FMM is unable to handle certain density profiles, such as those with prominent local extrema; without information on the direction of the gradient, the FMM will confuse local minima for local maxima. Anisotropic FMM solvers could address this limitation but are beyond the scope of this paper \cite{hassouna, sethian-and-vladimirsky}. Given these complexities, we recommend the continued use of synthetic profiles for analysis verification in future applications of this technique. 

    The precision of this analysis technique is limited by the quality of the boundary condition and the $\theta$ interpolation. Additional measurements could supplement AFR and enhance the analysis. For example, combining AFR with simultaneous interferometric imaging, as suggested by \textcite{angland}, would provide measurements of $\int n_e dz$ in the dilute plasma region that could serve as a boundary condition. Simultaneous AFR and interferometry measurements are already possible at OMEGA EP and have been demonstrated by \textcite{haberberger-inner-shell}. Improved interpolation and spatial resolution can be achieved by using multiple filters that measure different $\theta$ contours. \textcite{heuer-filters} has investigated the simultaneous use of two AFR filters. Alternatively, additional $\theta$ contours could be measured using multi-color AFR. Finally, combining AFR measurements with shadowgraphy, which has been performed in 1D by \textcite{schaeffer}, may improve spatial resolution and enable the detection of smaller-scale features such as shocks or instabilities. These improvements would expand the applicability of AFR and yield higher fidelity measurements of $\int n_e dz$. 

\section*{Acknowledgments}

     The authors thank the OMEGA EP team for their assistance while performing these experiments and the General Atomics team for target fabrication. The authors also thank F. Hernández Álvarez, N. Bohlsen, and S. Ivancic  for useful discussions. This work was supported by the Department of Energy under Grant No. DE-NA0004034. Experiment time was made possible by grant DE-NA0003613 provided by the National Laser User Facility.

\bibliography{bibliography}

\end{document}